\documentclass[]{aa}

\usepackage{natbib}

\usepackage{graphicx}
\usepackage[varg]{txfonts}

\def\Cau  {\ion{Ca}{i}} 
 
\def\Feu  {\ion{Fe}{i}}

\def\Liu  {\ion{Li}{i}}
\def\Siu  {\ion{Si}{i}}

\def\Srd  {\ion{Sr}{ii}}

\def\CHdou {$\rm ^{12}CH$}
\def\CHtre {$\rm ^{13}CH$}
\def\Teff  {$T_\mathrm{eff}$}
\def\logg  {$\log g$}
\def\loggf {$\log gf$}
\def\vt    {$\rm v_{t}$}
\def\kms   {$\rm km\,s^{-1}$}

\newcommand{\cobold}{{CO$^5$BOLD}}

\begin{document}

   \title{
Abundances in a sample of turnoff and subgiant stars 
in NGC 6121 (M4)
\thanks{Based on observations collected at the European Organisation for Astronomical Research in the Southern Hemisphere under ESO programme 085.D-0537(A)}
}
\author {
M. Spite\inst{1}\and
F. Spite\inst{1}\and
A. J. Gallagher\inst{1}\and
L. Monaco\inst{2}\and
P. Bonifacio\inst{1}\and
E. Caffau\inst{1}\and
S. Villanova\inst{3}
 }

\institute {
GEPI, Observatoire de Paris, PSL Research University, CNRS, Universit\'e Paris Diderot, Sorbonne Paris Cit\'e, Place Jules Janssen, 92195, Meudon
\and
Departamento de Ciencias Fisicas, Universidad Andres Bello, 220 Republica, Santiago, Chile
\and
Universidad de Concepci\'on, Departamento de Astronom\'ia, Casilla 160-C, Concepci\'on, Chile
}


\authorrunning{Spite et al.}

\titlerunning{Abundances in a sample of M4 turnoff stars}

  \abstract
{The stellar abundances observed in globular clusters show complex structures, currently not yet understood. 
}
{The aim of this work is to investigate the relations between the abundances of different elements in the globular cluster M4, selected for its uniform deficiency of iron, to explore the best models explaining the pattern of these observed abundances. 
Moreover, in turnoff stars, the abundances of the elements are not suspected to be affected by internal mixing.
}
{In M4, using low and moderate resolution spectra obtained for 91 turnoff (and subgiant) stars with the ESO FLAMES-Giraffe spectrograph, we have extended previous measurements of abundances (of Li, C and Na) to other elements (C, Si, Ca, Sr and Ba), using model atmosphere analysis. We have also studied the influence of the choice of the microturbulent velocity.
}
{Firstly, the  peculiar turnoff star found to be very Li-rich in a previous paper does not show any other abundance anomalies relative to the other turnoff stars in M4.
~Secondly, an anti-correlation between C and Na has been detected, the slope being significative at more than $3\sigma$. This relation between C and Na is in perfect agreement with the relation found in giant stars selected below the RGB bump.
~Thirdly, the strong enrichment of Si and of the neutron-capture elements Sr and Ba, already observed in the giants in M4, is confirmed.
~Finally, the relations between Li, C, Na, Sr and Ba constrain the enrichment processes of the observed stars.
}
{The abundances of the elements in the turnoff stars appear to be compatible with production processes by massive AGBs,  but are also compatible with 
the production of second generation elements (like Na) and low Li produced by, for example, fast rotating massive stars.}


\maketitle

%
\section{Introduction}
Globular clusters provide an opportunity to improve our understanding of the problems raised by stellar nucleosynthesis.
Spectroscopic and photometric observations indicate a complex structure of the stellar systems of these clusters;
 it is generally accepted that the observed abundances are the results of successive generations \citep[see for example][]{MarinoVP08,VillanovaG11,GrattonCB12,DoraziGA15,BragagliaCS15}, and also \citet{BastianCS15}.

Measured abundances of the elements are not easy to interpret, and some further analyses of stars in globular clusters are necessary. The M4 cluster is the nearest, enabling a detailed analysis which reveals some details in stars fainter than the stars in the giant branch. 
 
The M4 stars are known to have a very uniform abundance of iron \citep[for example][]{CarrettaBG09,MucciarelliSL11,MonacoVB12}, and a rather strong enhancement of the neutron capture elements, compared to the field stars with the same metallicity \citep{IvansSK99,YongKL08}. The red giant branch (RGB) of M4 shows, in the U versus U-B colour diagram, two different sequences \citep{MarinoVP08}.

 \citet{IvansSK99} and \citet{MonacoVB12} have shown in giant and turnoff stars respectively, that the abundance of Na shows a large spread. \citet{MonacoVB12} investigated the relation between the abundance of Na and Li. They found, for the stars hotter than 5880\,K, a mean lithium abundance A(Li) = 2.13 with a very shallow slope for the Li-Na anticorrelation (we note that below 5880\,K, Li is destroyed by internal stellar convection). Moreover, they detected a peculiar star with a very high Li abundance (A(Li) = 2.87 in M4-37934). 

The aim of this work is to add to the work of \citet{MonacoVB12} by ascertaining the abundances of some other elements, in particular the carbon abundance, in order to acquire a broader view of the abundance pattern of this cluster.

\section {Observational data} 
Observations were conducted at the Very Large Telescope (VLT) 
 (Paranal, Chile) between April and July 2010 using the LR2 setting from 400 to 456\,nm with a resolving power R = 6000, the HR12 setting from 583 to 614\,nm, and the HR15N setting from 666 to 679\,nm, both with a resolving power of about R=20000. The frames were processed using the FLAMES-GIRAFFE reduction pipeline. More information can be found in \citet{MonacoVB12}. The spectroscopic data are available through the Giraffe archive at Paris Observatory ({\tt http://giraffe-archive.obspm.fr/}). 

A total of 91 stars in M4 are discussed in the present paper.
These stars span a very small interval of the HR diagram. Among them, 71 are turnoff stars with temperatures higher than 5600\,K  and gravities between 4.1 and 3.8. Ten other stars are subgiants and have a temperature lower than 5500\,K \citep[see the Fig. 1, in ][]{MonacoVB12},  and gravities between 3.5 and 3.7.

\section {Spectral analysis and abundance measurements} 
Unlike \citet{MonacoVB12} who used ATLAS models with overshooting and
the MOOG code,  in our analysis we used OSMARCS model atmospheres \citep {GustafssonEE08} together   with  the {\sf turbospectrum} synthesis code \citep{AlvarezP98,Plez12}. In this code, the collisional broadening by neutral hydrogen is generally computed following the theory developed by \citet{AnsteeO91,BarklemPO2000,BarklemO2000}.\\
 For each star, the model parameters have been adopted from \citet{MonacoVB12}: the temperature is mainly based on photometry, the gravities on theoretical isochrones, and the microturbulent velocity has been chosen to obtain an abundance of \Feu~  independent of the equivalent width of the lines. It has been found \citep{MonacoVB12} that this microturbulence velocity decreases  with the evolution of the star, from 1.7 \kms~ for the turnoff stars to 1.0 \kms~ for the subgiants. 

\citet{MonacoVB12}  found $\rm [Fe/H] = -1.31~dex \pm 0.03$ for the turnoff stars and $\rm [Fe/H] = -1.17~dex \pm 0.03$ for the subgiants. 
We compared the iron abundance found with the two methods (OSMARCS models + turbospectrum and ATLAS with overshooting + MOOG) and the same model parameters in a subsample of 18 stars, and found a slight systematic difference of $\rm +0.11 \pm 0.03$. 
Applying this correction to the \citet{MonacoVB12} values, we found for the turnoff stars of the sample $\rm [Fe/H] = -1.20 ~dex \pm 0.06$ which is in good agreement with the values found by \citet{YongKL08} ([Fe/H] =--1.2), 
\citet{CarrettaBG09} ([Fe/H] = --1.2), \citet{IvansSK99} ([Fe/H] = --1.18), and \citet{MucciarelliSL11} ([Fe/H] = --1.10).
However, as in \citet{MonacoVB12}, we found that the subgiants of the sample are systematically more metal-rich ($\rm [Fe/H] = -1.10 ~dex \pm 0.06$) than the turnoff stars.\\

A decrease in the microturbulence velocity between turnoff and subgiant stars is a priori, unexpected. Generally, a slight increase of the microturbulence velocity (0.2 \kms) is observed when a star evolves from turnoff to subgiant  \citep[see e.g.][]{EdvardssonAG93}. Since this decrease  was based on a rather small number of iron lines we decided to investigate the consequences of a change of the microturbulence velocity adopting, as an alternative, \vt = 1.3 \kms~ for all the stars (this value remains compatible with the observed spectra).
In this case we found that the iron abundance in all the stars, is uniform and equal to $\rm [Fe/H] = -1.16 ~dex \pm 0.06$.

It is important to note that in \citet{MonacoVB12}, the lithium abundance has been directly derived from the equivalent width of the \Liu~ resonance doublet at 670.8 nm using the Sbordone et al. (2010) formula B.1 based on 3D models and  non local thermodynamical equilibrium (NLTE) computations. As a consequence, the lithium abundance is not affected by the difference of the model grids used in \citet{MonacoVB12} nor by the microturbulence velocity. In the discussion we adopted the lithium abundances given by \citet{MonacoVB12} in their Table A1.
For all the other elements the abundances have been computed with the microturbulence velocity given in \citet{MonacoVB12} (option A), and with \vt = 1.3 \kms (option B).
The impact of a change in the microturbulent velocity depends on the equivalent widths of the lines. It has practically no effect on weak or very strong lines but is maximum around 100\,m\AA. The effect is often the same on the element ``X'' and on the iron abundance, in this case [X/Fe] does not change.

\subsection {Carbon abundance} 
\begin{figure}[h]
\resizebox{\hsize}{!}                   
{\includegraphics {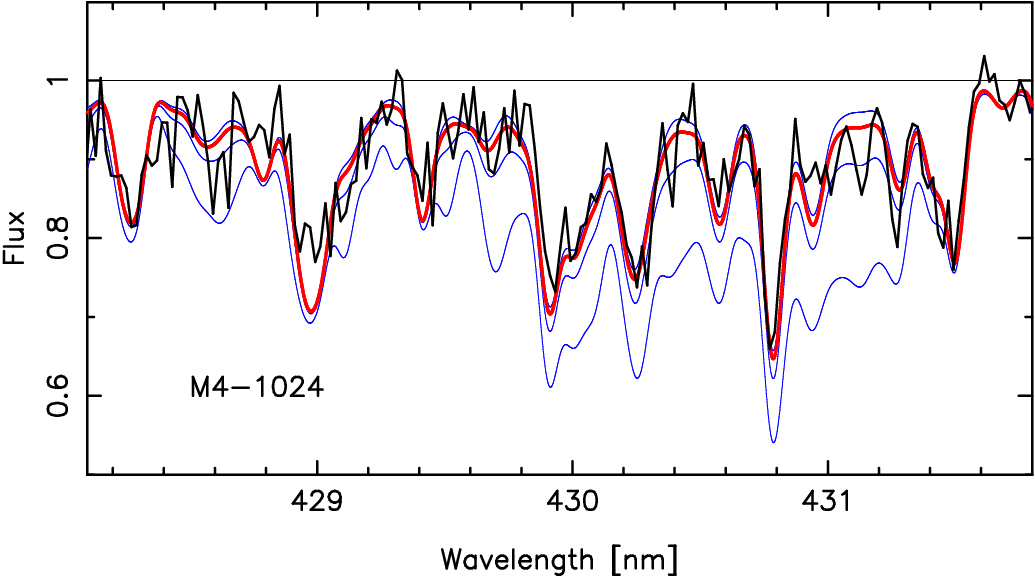}}
\resizebox{\hsize}{!}
{\includegraphics {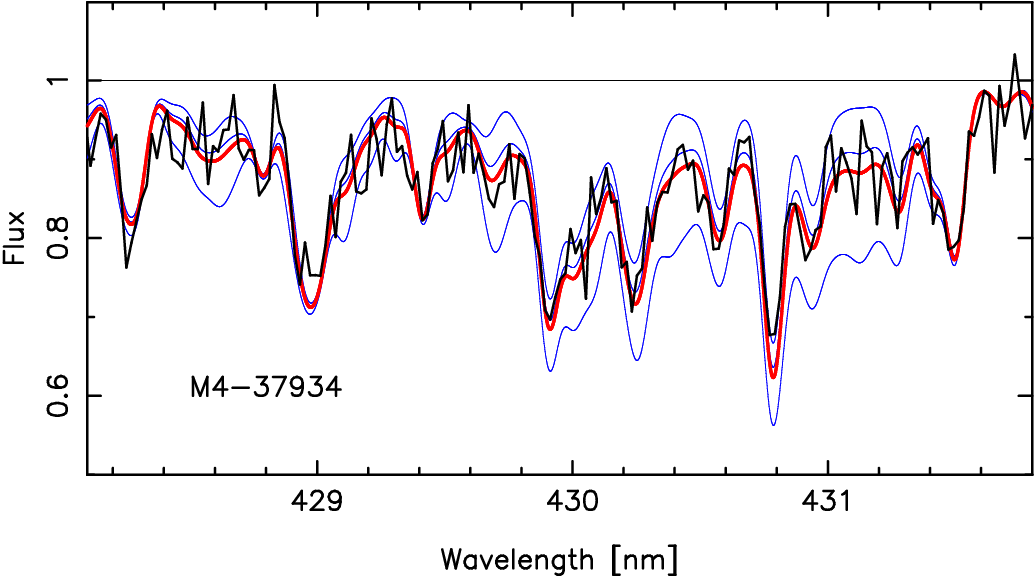}}
\caption[]{Observed profile of the CH band in M4-1024 and M4-37934~(black line). In both cases synthetic profiles (blue thin lines) have been computed with A(C)= 6.4 6.9 7.4. The thick red line represent the best fit (1D computations) A(C)=6.6 for M4-1024 and A(C)=7.0 for M4-37934~.}
\label {CHband}
\end{figure}

The carbon abundance has been computed by fitting the computed profile of the CH G-band ($A^{2} \Delta - X^{2} \Pi $) to the observed spectrum between 426 and 431.6\,nm. Line lists for \CHdou ~and \CHtre~ \citep{MasseronPVE14} were included in the synthesis.
The carbon abundances were computed independently adopting the microturbulence velocity given by \citet{MonacoVB12}, and  also \vt = 1.3 \kms.  The  CH band is not very sensitive to a change of the microturbulence velocity and the difference in A(C) between the two options never exceeds 0.1 dex, it is about +0.02 dex for the subgiants and --0.01 dex for the turnoff stars.
In Fig. \ref{CHband} we present the fit of the spectra for two typical stars in M4 computed with the microturbulence velocity given by \citet{MonacoVB12}.

The CH band is formed in a region of the atmosphere close to the surface where the effects of stellar convection can be important. Three dimensional (3D) dynamic model atmospheres \citep[CIFIST models based on the \cobold ~code:][]{LudwigCS09,FreytagSL12} were used to compute spectral synthesis and test the role of the granulation in the formation of the G band 
\citep{GallagherCB16}. The models they investigated have atmospheric parameters of turnoff stars, close to the stars of M4 analysed here. We were able to
interpolate corrections from their Table 2 to derive a 3D correction to apply to the C abundance derived here.
This correction $\rm A(C)_{3D} - A(C)_{1D}$, is very small: approx. +0.03\,dex for the M4 turnoff stars and +0.05\,dex for the subgiants.

\subsection {Sodium abundance} \label{Na-ab}
Using the OSMARCS model atmospheres and turbospectrum, we computed the profiles of the sodium D lines. Applying the NLTE correction computed by S. Andrievsky and S. Korotin \citep[as in][]{MonacoVB12}, we found a very small systematic difference with the values published by \citet{MonacoVB12}: $\rm \Delta_{A(Na)} =-0.06 \,dex \pm 0.06$  but the scatter of [Na/Fe] remains the same. 

The strong sodium D lines are not very sensitive to a change of the microturbulence velocity and as a consequence a change of the microturbulent velocity does not change the scatter of [Na/Fe].\\ 
We estimate that the error in the A(Na) measurement is about 0.1 \,dex.\\It is interesting to note that also in metal-poor turnoff field stars, the abundance of Na at this metallicity is scattered \citep[see e.g.][]{GehrenSZ06}.

\subsection {Silicon abundance} \label{Si-ab}
The silicon abundance has been deduced from only the \Siu ~line at 594.855 nm (Table \ref{tablines}). The resolving power of the spectrum in this region is close to 20000 and the S/N per pixel, is between 55 and 85. The equivalent width (EW) of the \Siu ~line is between 25 and 70 m\AA~  and thus the silicon abundance is sensitive to the microturbulence velocity. 
The expected uncertainty in the measured equivalent width can be estimated by the Cayrel formula \citep{Cayrel88}:
\begin{equation}
\sigma_{EW}=  \frac{1.5}{S/N} \sqrt{FWHM * \delta x}
\end{equation}
where S/N is the signal-to-noise ratio per pixel, FWHM the full width of the line at half maximum, and $\delta x$ the pixel size. For a typical S/N ratio of 70 per pixel the predicted accuracy $\sigma_{EW}$ is about 3~m\AA.
However this formula neglects the uncertainty on the continuum placement. After varying slightly the continum level within the range allowed by the S/N of the spectra,  we estimated that the accuracy in the measurement of the equivalent width of the \Siu ~line is about 6~m\AA. This uncertainty corresponds to an uncertainty of about 0.1 \,dex in [Si/Fe] but it relies on only one \loggf ~value classified as quality C ($ \sigma(gf) \leq 25\%$), in the NIST data base \citep{NIST2012}.

\begin{table}
\begin{center}    
\caption[]{
Main characteristics of the atomic lines used in the present analysis.
R is the Resolving power of the spectrum used for the measurement}
\label{tablines}
\begin{tabular}{lccccccc}
\hline\hline
Elem  & wavelength & $\rm \chi_{ex}$ &  log {\it gf} & {\it R} & Remark \\
      & (nm)       &    (eV)         \\ 
\hline
Si I  &  594.855  & 8.50  &  -1.23 & 20000 \\
~~~   \\
Ca I  &  610.272  & 1.88  &  -0.79 & 20000 \\
Ca I  &  612.221  & 1.89  &  -0.32 & 20000 \\
Ca I  &  649.378  & 2.52  &  -0.11 & 20000 \\
Ca I  &  671.768  & 2.71  &  -0.52 & 20000 \\
~~~   \\      
Ni I  &  676.772  & 1.83  &  -2.17 & 20000 \\
~~~   \\
Sr II &  407.771  & 0.00  &  +0.17 &  6000 \\
~~~   \\
Ba II &  455.40   & 0.00  &  +0.17 & 6000  & hfs(1)\\
Ba II &  585.37   & 0.60  &  -1.01 & 20000 & hfs(1)\\ 
Ba II &  649.69   & 0.60  &  -0.41 & 20000 & hfs(2)\\     
\hline
\end{tabular}
\tablefoot{hfs(1): hyperfine structure from \citet{McWillSS95}. hfs(2): hyperfine structure from \citet{MashonkinaGB99}.}
\end{center}
\end{table}

\subsection {Calcium abundance} \label{Ca-ab}
The calcium abundance was deduced from the fit of the profile of four Ca lines (Table \ref{tablines}). For each star, the accuracy of the Ca abundance was estimated from the standard deviation of these four abundances around the mean value, the mean standard deviation is close to $\rm \approx 0.11\, dex$.       
The equivalent widths of the four \Cau~ lines are close to 100\,m\AA~ and thus are sensitive to the microturbulence velocity.
The calcium lines are also sensitive to NLTE effects. The NLTE corrections have been computed by S. Andrievsky and are available in \citet{SpiteAS12} or via {\tt \footnotesize http://cdsarc.u-strasbg.fr/viz-bin/qcat?J/A+A/541/A143}.
For the Ca lines used in the present paper this NLTE correction is close to --0.11 dex.

\subsection {Nickel abundance} \label{Ni-ab}
The nickel abundance was estimated from the equivalent width of only one line (Table \ref{tablines}) as was done for silicon. The equivalent width of this Ni line varies from star to star, between 14 and 70 m\AA,  the strongest lines are thus sensitive to a change of the microturbulence velocity. We took into account only the Ni lines that had an equivalent width larger than 20 m\AA. In this region of the spectrum, the resolving power of the spectra is close to 20000 and the S/N ratio per pixel is between 50 and 100, with a typical value of about 70. From the Cayrel formula (see section \ref{Si-ab}) we found $\rm \sigma_{EW} \approx 3~m\AA$, and 6~m\AA ~taking into account the uncertainty on the position of the continuum. With a typical equivalent width of 30 m\AA ~this value corresponds to an uncertainty in [Ni/Fe] of about 0.14 \,dex. Moreover the Ni abundance relies on only one \loggf ~value classified as quality D ($ \sigma(gf) \leq 50\%$) in the NIST data base \citep{NIST2012}. 

\subsection {Strontium abundance} \label{Sr-ab}
\begin{figure}[h]
\resizebox{\hsize}{!}                   
{\includegraphics {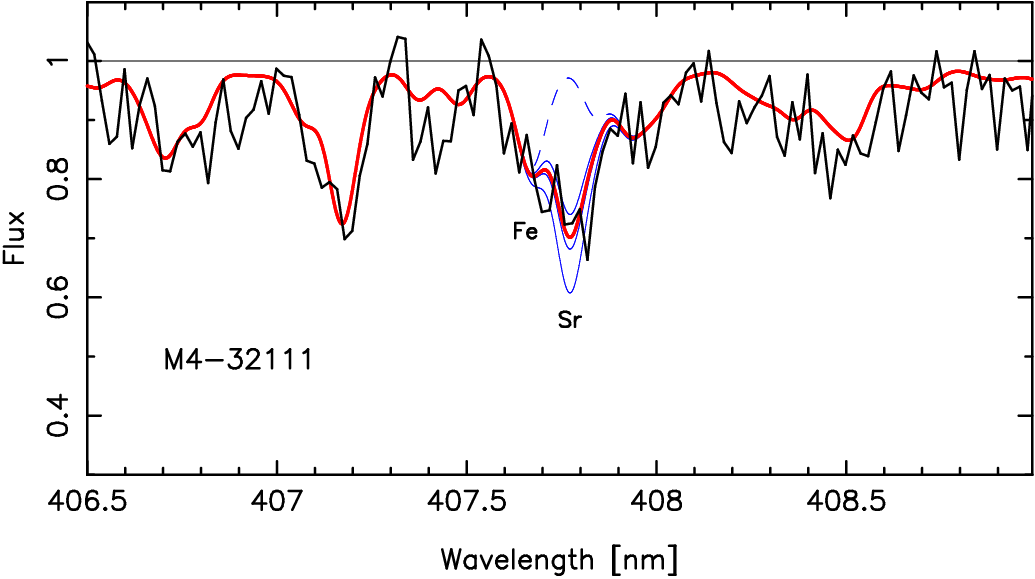}}
\caption[]{Observed profile of the feature around the \Srd ~line (black line). Three synthetic profiles (blue thin lines) have been computed with abundances A(Sr) = 1.9, 2.2, 2.5 and one with no strontium (dashed blue line). The thick red line represent the best fit  (LTE computations) with A(Sr)=2.1 for the star M4-32111.}
\label {Sr2line}
\end{figure}

The strontium abundance has been estimated from the fit of the observed spectrum around the Sr line at 407.77\,nm with a synthetic spectrum computed with {\tt turbospectrum}.
The observed spectrum in the region of the Sr line only has a resolving power R = 6000, an example of the fit is given in Fig. \ref{Sr2line}. The uncertainty of the measurement of [Sr/Fe] is estimated to be as large as 0.25\,dex.

\subsection {Barium abundance} \label{Ba-ab}
We derived the abundance of Ba from the fitting of three Ba features taking into account the hyperfine structure of the lines (Table \ref{tablines}). For twelve stars the standard deviation is equal or larger than 0.30. These stars have not been taken into consideration. Moreover, for nine other stars only the "blue" low resolution spectra (setting LR2) were available and thus the Ba abundance has been estimated only from the resonance line at 455.4\,nm and we estimate that for these stars the stochastic uncertainty can reach 0.25 \,dex. For the other 70 stars the mean standard deviation is equal to 0.15 \,dex. 

\subsection {Error budget}
Uncertainties in the abundance measurements include those related to the adopted oscillator strengths, to the equivalent widths or profile measurements and to the adopted stellar parameters.
The error on the oscillator strengths is difficult to assess and can be particularly important when the element is represented by a very small number of lines. However, since this error is the same for all the stars, it can produce only a general shift in the abundance of the element (the same for all the stars).\\
The adopted measurement errors are given for each element in sections \ref{Na-ab} to \ref{Ba-ab}.\\

The computations of the error linked to our choice of the stellar parameters (Table \ref{errors}) have been done for a star with \Teff~= 5950\,K and \logg~= 4.0 (model A).
A change of --100\,K of the temperature corresponds to model B.
Since gravity has been estimated with the aid of theoretical isochrones \citep{MonacoVB12}, a change of temperature of --100\,K would induce an automatic change of \logg~ of --0.1\,dex (model C). 
A comparison of the results obtained with models A and B, and B and C, allows an estimation of the error induced by a variation of \Teff~ and \logg~ alone.
A comparison of the results obtained with models C and A shows the global consequences of a change of --100\,K in the temperature of the stars: it does not significantly affect the ratios [X/Fe] when atomic lines are used (for Na, Si, Ca, Ni, Sr, Ba). But [C/Fe], based on the study of the CH band, decreases by 0.13\,dex.

For each element X, the total uncertainty on the relative abundance [X/Fe] is computed as the quadratic sum of the stochastic and systematic errors. The total uncertainty is dominated by the measurement errors, since errors in the model parameters largely cancel each other 
out in the measured ratios of the elements.

\begin{table}
\begin{center}    
\caption[]{
Abundances uncertainties linked to stellar parameters}
\label{errors}
\begin{tabular}{lccccccc}
\hline\hline
Elem  & $\rm \Delta(B-A)$   &$\rm \Delta(C-B)$& $\rm \Delta(C-A)$\\
\hline
$\rm[Fe/H] $&   -0.10&  +0.03  & -0.07     \\
$\rm[C/Fe] $&   -0.17&  +0.04  & -0.13     \\
$\rm[Na/Fe]$&   +0.03&  -0.03  & +0.00     \\
$\rm[Si/Fe]$&   +0.04&  -0.01  & +0.03     \\ 
$\rm[Ca/Fe]$&   +0.04&  -0.01  & +0.03     \\
$\rm[Ni/Fe]$&   -0.03&  +0.01  & -0.02     \\ 
$\rm[Sr/Fe]$&   -0.01&  -0.01  & -0.02     \\
$\rm[Ba/Fe]$&   +0.03&  -0.03  & +0.00     \\    
\hline
\end{tabular}
\tablefoot{\\
        A: \Teff=5950\,K, \logg=4.0, \vt=1.7 \kms. \\
        B: \Teff=5850\,K, \logg=4.0, \vt=1.7 \kms. \\
        C: \Teff=5850\,K, \logg=3.9, \vt=1.7 \kms}
\end{center}
\end{table}

\begin{figure}[h]
\resizebox{\hsize}{!}                   
{\includegraphics {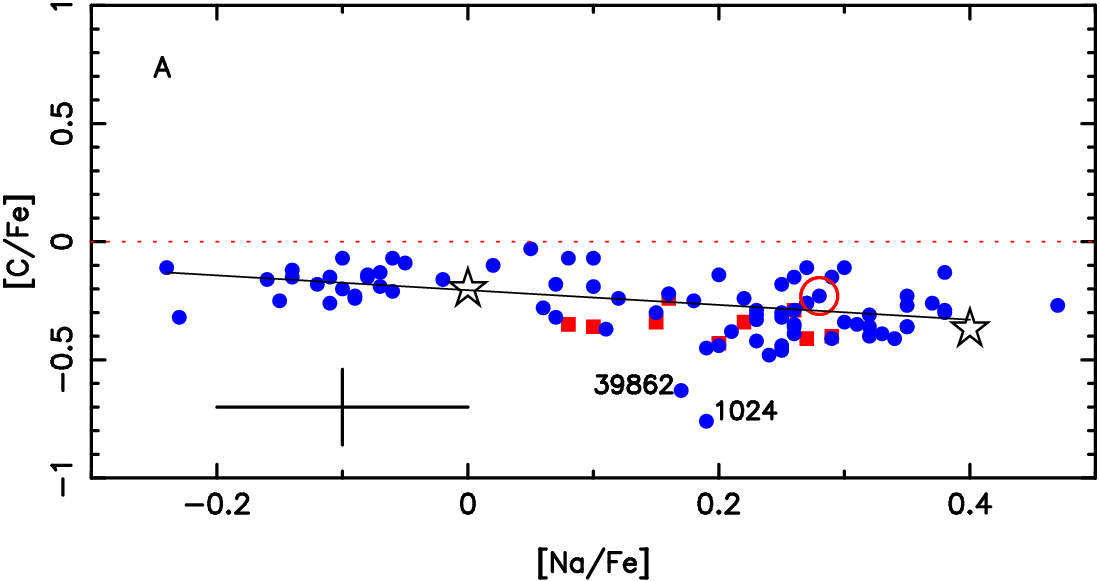}}
\resizebox{\hsize}{!}                   
{\includegraphics {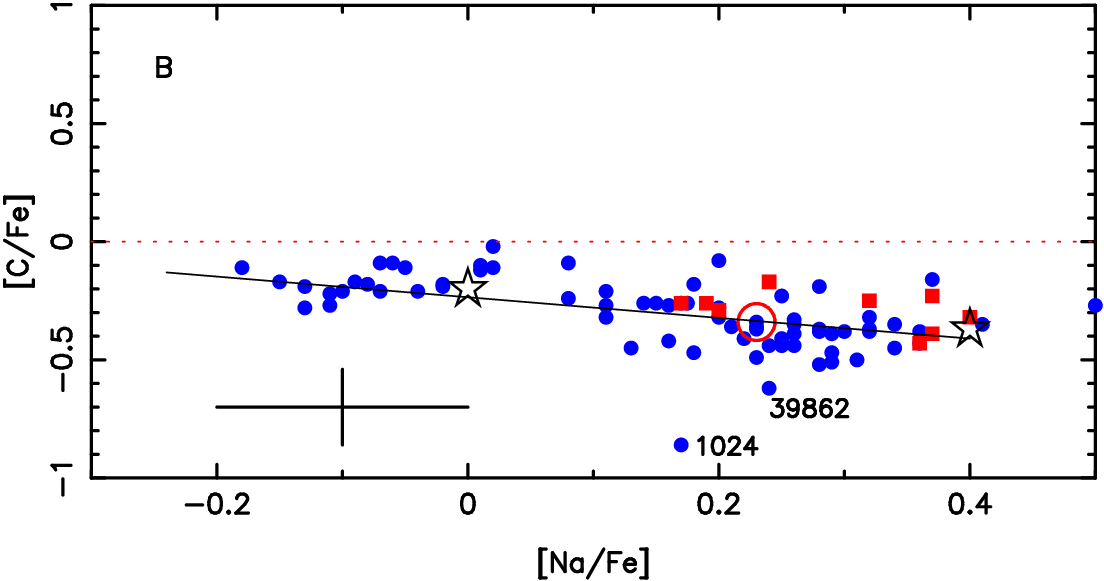}}
\caption[]{Relation between [C/Fe] and [Na/Fe] computed, A) with the model parameters \Teff~ \logg~ and \vt~ given in \citet{MonacoVB12}, and B) with the same parameters but \vt~=1.3\kms. The red squares represent the subgiant stars which have a temperature below 5600\,K. The blue dots are for the turnoff stars with 
$\rm 5600\,K < T_{eff} < 5980\,K $. The very Li-rich star M4-37934 is surrounded by a red circle. A typical value of the total error is represented by the cross at the lower left corner of the figure. The mean values obtained for the two groups of RGB stars in M4 by \citet{VillanovaG11}: C-rich N-poor stars and the C-poor N-rich stars are shown by black open stars symbols.
}
\label {CFe}
\end{figure}

\section {Discussion}
Tables \ref{abund1} and \ref{abund2} tabulate the abundances of lithium, carbon, sodium, silicon, calcium, strontium and barium relative to iron in our sample of turnoff and subgiant stars, adopting respectively the microturbulence velocity given in \citet{MonacoVB12} (option A) and \vt~= 1.3 \kms (option B). 
In this last case we adopted for all the stars  [Fe/H]=--1.16\,dex since the measurement errors exceeded the measured scatter.
For several elements the effect of a variation of the microturbulence velocity on the abundance is about the same as it is for Fe and thus [X/Fe] is little affected.

For Li and Na the abundances given in the tables take into account the NLTE effects. The sodium abundance has been determined from the resonance lines and thus the NLTE correction is large and it is different from star to star since it depends on the equivalent width of the Na lines. In order to compare with the Na abundances in the literature,                                                                                                                                                                                                                                                        generally estimated from the secondary lines, that are less affected by NLTE effects, this correction has to be carefully taken into account. 
Since generally in the literature the Ca and Ba abundances are not corrected for NLTE effects, we kept in the tables the LTE Ca and Ba abundances for an easier comparison. When possible, an estimation of the NLTE effects is given in the discussion.

\begin{table*}
\begin{center}    
\caption[]{
Abundances measured in our sample of M4 stars computed with the model parameters \Teff~ \logg~ and \vt~ given by \citet{MonacoVB12}. The full table is available at the CDS.}
\label{abund1}
\begin{tabular}{l@{~~}c@{~~}c@{~}c@{~}c@{~~}c@{~~}c@{~~}c@{~~}c@{~~}c@{~~}c@{~~}c@{~~}c@{~~}c@{}c}

\hline\hline
\multicolumn{5}{c}{adopted model}&NLTE&NLTE&NLTE,3D&3D-LTE&1D-LTE&1D-LTE&1D-LTE&1D-LTE&1D-LTE\\
\cline{1-5}
ID&Teff&logg&vt&[Fe/H]&[Na/Fe]&Na corr&A(Li)&[C/Fe]&[Si/Fe]&[Ca/Fe]&[Ni/Fe]&[Sr/Fe]&[Ba/Fe]\\
\hline
 506&5930&4.0&1.7&-1.24& 0.23&-0.18& 2.10& -0.33& 0.34& 0.08& 0.25& 0.12& 0.17\\
1024&5920&4.0&1.7&-1.21& 0.19&-0.17& 2.10& -0.79& 0.51& 0.08& 0.20& 0.39& 0.39\\
7746&5910&4.0&1.7&-1.21& 0.32&-0.14& 2.21& -0.36& 0.36& 0.12& 0.18& 0.29& 0.29\\
8029&5920&4.0&1.7&-1.22& 0.26&-0.16& 2.13& -0.35& 0.40& 0.12&  -  & 0.15& 0.35\\
8332&5930&4.0&1.7&-1.23& 0.25&-0.15& 2.03& -0.44& 0.37& 0.11&  -  & 0.46& 0.35\\
8405&5900&4.0&1.7&-1.23& 0.12&-0.17& 2.30& -0.24& 0.49& 0.32&  -  & 0.46& 0.41\\
8784&5950&4.0&1.7&-1.22&-0.14&-0.24& 2.17& -0.15& 0.59& 0.21& 0.21& 0.20& 0.36\\
... \\
\hline
\end{tabular}
\end{center}
\end{table*}

\begin{table*}
\begin{center}    
\caption[]{
Abundances measured in our sample of M4 stars computed with the model parameters \Teff~ \logg~ given by \citet{MonacoVB12} but with \vt =1.3 \kms . The full table is available at the CDS.}
\label{abund2}
\begin{tabular}{l@{~~}c@{~~}c@{~}c@{~}c@{~~}c@{~~}c@{~~}c@{~~}c@{~~}c@{~~}c@{~~}c@{~~}c@{~~}c@{}c}
\hline\hline
\multicolumn{5}{c}{adopted model}&NLTE&NLTE&NLTE,3D&3D-LTE&1D-LTE&1D-LTE&1D-LTE&1D-LTE&1D-LTE\\
\cline{1-5}
ID&Teff&logg&vt&[Fe/H]&[Na/Fe]&Na corr&A(Li)&[C/Fe]&[Si/Fe]&[Ca/Fe]&[Ni/Fe]&[Sr/Fe]&[Ba/Fe]\\
\hline
 506&5930&4.0&1.3&-1.16& 0.16& -0.18&2.10 &-0.42& 0.27& 0.07& 0.20& 0.20& 0.28\\
1024&5920&4.0&1.3&-1.16& 0.17& -0.17&2.10 &-0.86& 0.49& 0.09& 0.18& 0.39& 0.53\\
7746&5910&4.0&1.3&-1.16& 0.32& -0.14&2.21 &-0.38& 0.33& 0.14& 0.16& 0.35& 0.43\\
8029&5920&4.0&1.3&-1.16& 0.23& -0.16&2.13 &-0.37& 0.36& 0.12&-    & 0.33& 0.47\\
8332&5930&4.0&1.3&-1.16& 0.26& -0.15&2.03 &-0.44& 0.32& 0.11&-    & 0.47& 0.46\\
8405&5900&4.0&1.3&-1.16& 0.11& -0.17&2.30 &-0.27& 0.44& 0.32&-    & 0.36& 0.52\\
8784&5950&4.0&1.3&-1.16&-0.15& -0.24&2.17 &-0.17& 0.55& 0.21& 0.18& 0.39& 0.48\\
... \\
\hline
\end{tabular}
\end{center}
\end{table*}

\subsection{Relation between the abundances of C and Na.}
        
In globular clusters the abundances are often measured in bright giants. But in giant stars the carbon abundance may be affected by mixing with deep layers processed by the CNO cycle. 
Carbon abundances in turnoff stars, however, should be an indisputable measurement of the C abundance in the gas cloud that formed the star.

In Fig. \ref{CFe} we present the variation of the carbon abundance [C/Fe] in the M4 turnoff stars and the subgiants as a function of [Na/Fe] with the two different choices of the microturbulence velocity (options A and B). This figure includes the NLTE correction for the Na abundance and the weak 3D correction for the carbon abundance (see Tables \ref{abund1} and \ref{abund2}).
Two turnoff stars appear to be very carbon-poor (and slightly Na-rich): M4-1024 and M4-39862. 

If we ignore these two C-poor stars, an anti-correlation between C and Na is observed in the turnoff stars (slope $-0.32$, correlation factor $-0.51$ when the microturbulence velocity given in \citet{MonacoVB12} is adopted,  and a slope $-0.43$, with a correlation factor $-0.59$ when \vt~=1.3 \kms~is adopted). 
A non parametric test (Kendall's tau) shows that the probability of correlation is better than 99.98\%.

On Fig. \ref{CFe}A the subgiant stars seem to have, as a mean, a carbon abundance systematically 0.1 dex lower than the turnoff stars with the same value of [Na/Fe] (if we ignore the two C-poor turnoff stars).
However this effect disappears (Fig. \ref{CFe}B) when a microturbulent velocity \vt~=1.3 \kms~ is adopted for subgiant and turnoff stars.\\
In these figures the subgiant stars all have a positive value of [Na/Fe], but this effect is spurious. 
In our sample, there is at least 
one subgiant (M4-33548) which has a lower value of [Na/Fe]: $-0.06\pm0.15$, unfortunately we could not measure the carbon abundance of this star and thus it does not appear on Figs. \ref{CFe}A and \ref{CFe}B (but it is visible on the subsequent figures).

 A sample of RGB stars located below the RGB bump, has been studied in M4 by \citet{VillanovaG11}. They found a correlation between [C/Fe] and [Na/Fe], interpreted as the result of the existence of two different populations.
Their sample is divided in two subsamples: the first one with the C-rich N-poor stars and the second one with the C-poor N-rich stars. In Fig. \ref{CFe}A and \ref{CFe}B we plotted the mean Na abundance of these two groups of giants as a function of the mean value of [C/Fe] (the black open stars symbols). The agreement between turnoff stars and RGB stars is excellent.

The position in the diagram of the very Li-rich turnoff star M4-37934 \citep{MonacoVB12}, is surrounded by a red circle in Fig. \ref{CFe}, this star has a normal carbon abundance compared to the other turnoff stars with the same value of [Na/Fe].

\subsection{Abundances of $\alpha$-elements: [Si/Fe], [Ca/Fe] vs. [Na/Fe].}
\begin{figure}[h]
\resizebox{\hsize}{!}                   
{\includegraphics {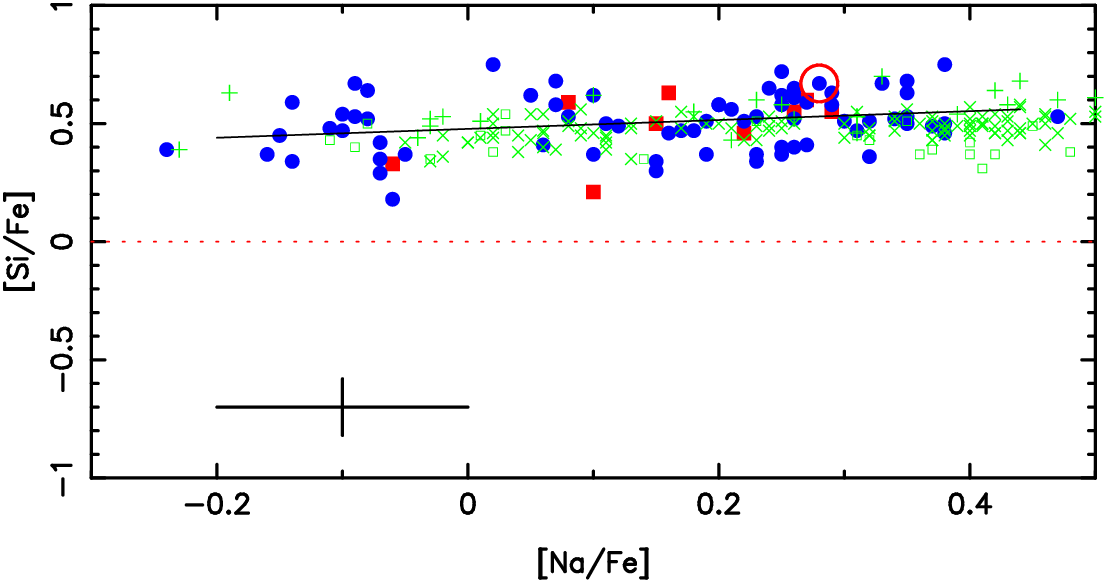}}
\resizebox{\hsize}{!}                   
{\includegraphics {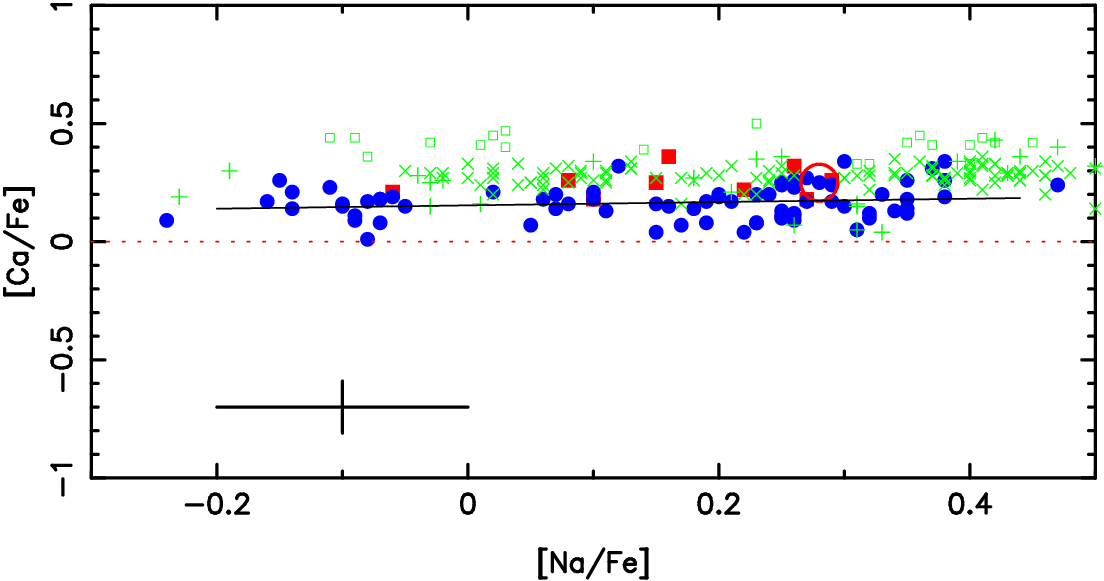}}
\caption[]{Relations between the abundances of the $\alpha$ elements Si (upper panel) and Ca (lower panel) vs. [Na/Fe]. The symbols  are as in Fig. \ref{CFe}. In these two figures we compare the abundance measured in the turnoff stars to the abundances measured by \citet{IvansSK99}, \citet{MarinoVP08} and \citet{VillanovaG11} in samples of bright giants (respectively green $+$, $\times$ and $\square$).
The $\alpha$-elements Si and Ca are overabundant relative to iron.}
\label {alfaFe}
\end{figure}

The relations of [Si/Fe] and [Ca/Fe] versus [Na/Fe] are displayed in Fig. \ref{alfaFe} for the microturbulence velocity of \citet{MonacoVB12}. Silicon and Calcium are overabundant relative to iron in M4 turnoff and subgiant stars. The mean overabundance of silicon is [Si/Fe] = +0.51 \,dex if the \vt~ of \citet{MonacoVB12} is adopted, and  [Si/Fe] = +0.48 if \vt~ = 1.3 \kms, but since this measurement is based on only one silicon line with a rather uncertain \loggf ~value, this result could be affected by a systematic error. However, \citet{VillanovaG11}, \citet{YongKL08}, \citet{MarinoVP08} and \citet{IvansSK99} also found a similar high abundance of silicon based on more Si lines ([Si/Fe]=+0.43, [Si/Fe]=+0.58, [Si/Fe]=+0.48 and +0.55), in the bright giants of M4. 

Contrary to this, (with the two different values of \vt), we found a rather low mean overabundance of Ca: [Ca/Fe]=+0.18  close to the mean value found by \citet{IvansSK99}: [Ca/Fe]=0.26, or \citet{MarinoVP08}: [Ca/Fe]=0.28, while \citet{YongKL08} and \citet{VillanovaG11} found in M4 [Ca/Fe]=+0.41. 

There is no trend of [Ca/Fe] vs [Na/Fe]. A positive correlation between [Si/Fe] and [Na/Fe] is possible, but there is a large scatter in [Si/Fe] and the slope is not significant.

The values of [Si/Fe] and [Ca/Fe] in the Very Li-rich star M4-37934, are similar to the values found for the other M4 turnoff stars.
 
\subsection{Ni abundances vs. [Na/Fe].}
\begin{figure}[h]
\resizebox{\hsize}{!}                   
{\includegraphics {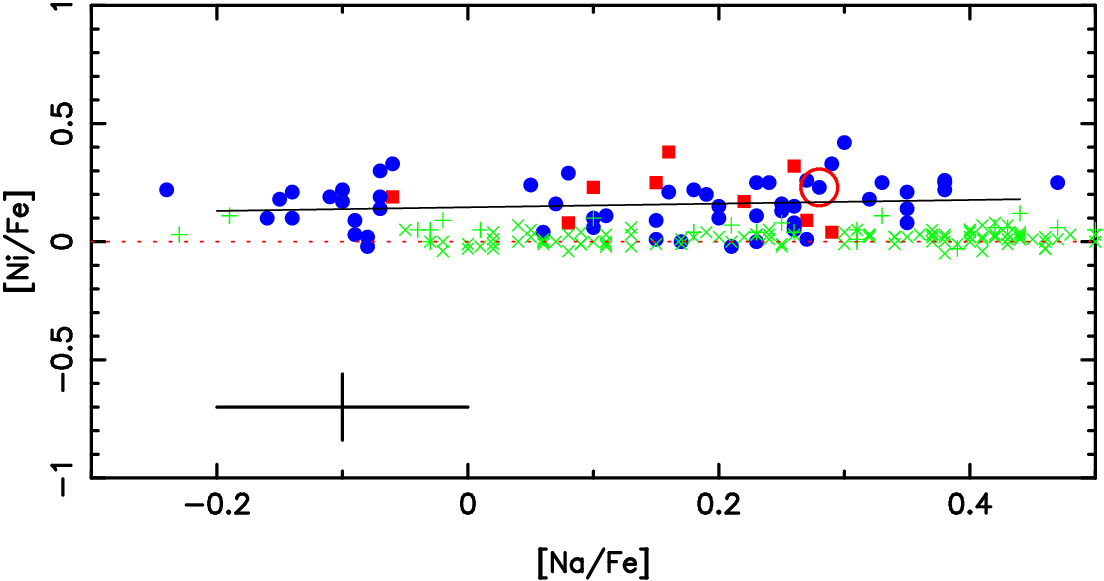}}
\caption[]{Relation between the abundance of the iron peak element [Ni/Fe] vs. [Na/Fe]. The symbols are the same as in Fig. \ref{CFe}. 
}
\label {NiFe}
\end{figure}

We found a slight overabundance of nickel in the turnoff and subgiant stars in M4: $\rm [Ni/Fe]=+0.16 \pm 0.10$ if for \vt ~the value given by \citet{MonacoVB12} is adopted and $\rm [Ni/Fe]=0.14 \pm 0.10$ if \vt = 1.3 \kms. There is no significant trend with [Na/Fe] (Fig. \ref{NiFe}). Taking into account that our determination of the Ni abundance is based on only one line with a rather uncertain \loggf~ (see section \ref{Ni-ab}), this mean value is in fair agreement with the values found in giants by \citet{IvansSK99} [Ni/Fe]=+0.05; \citet{MarinoVP08}  [Ni/Fe]=+0.02; \citet{VillanovaG11}  [Ni/Fe]=--0.01. It is in very good agreement with the value found by \citet{YongKL08}: [Ni/Fe]=+0.12 (these last data could not be plotted on the Fig. \ref{NiFe} because the sodium abundance has not been measured in the sample of \citet{YongKL08}). 

\subsection{Heavy elements abundances: Sr and Ba vs. [Na/Fe]}
\begin{figure}[h]
\resizebox{\hsize}{!}                   
{\includegraphics {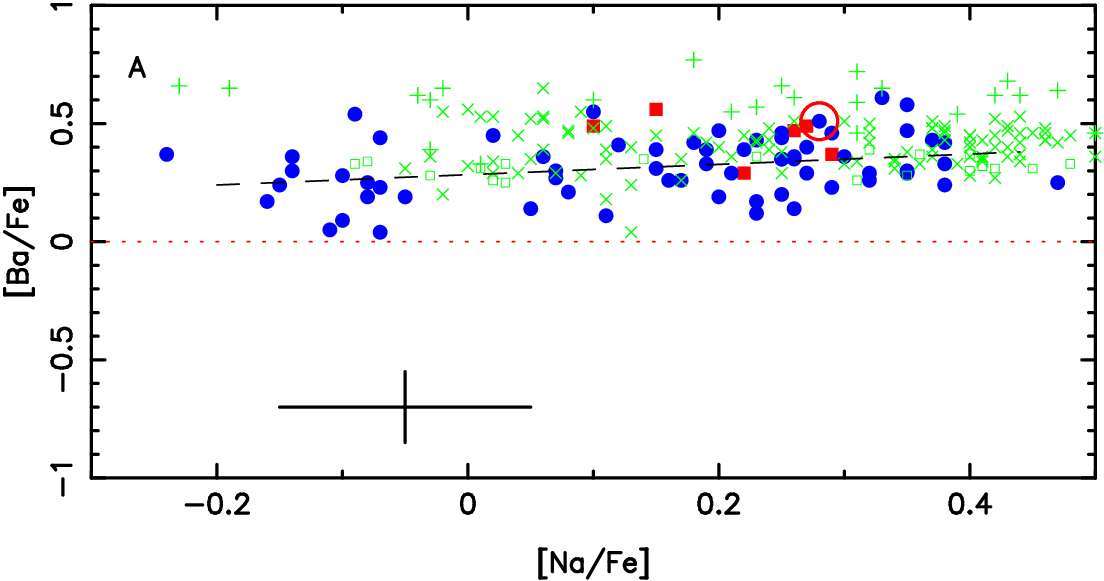}}
\resizebox{\hsize}{!}                   
{\includegraphics {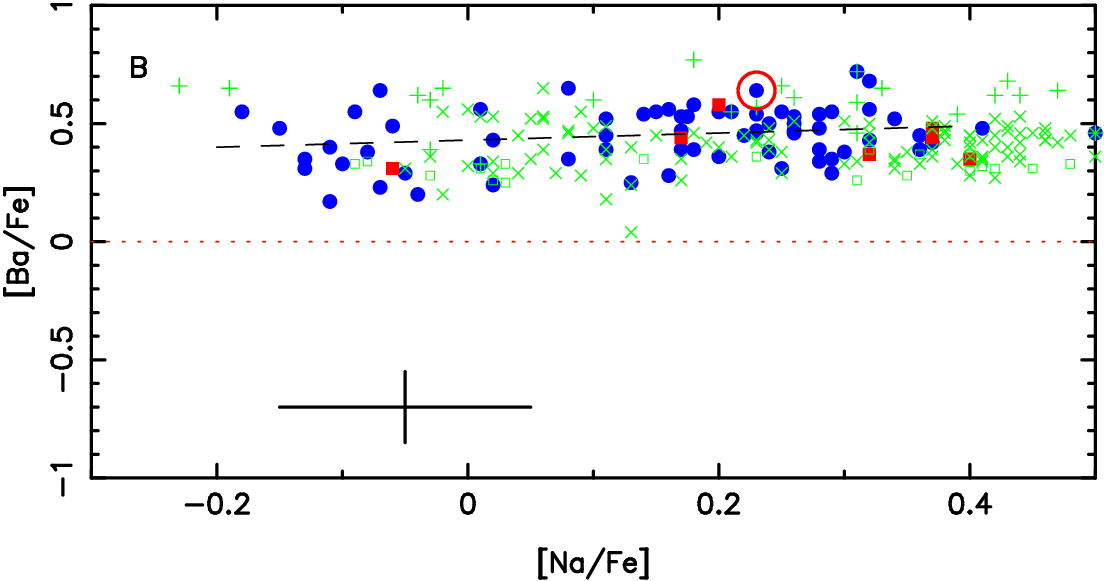}}
\caption[]{ Relation between [Ba/Fe] and [Na/Fe] computed with ~~~~~~~~~~~~~~~~~~~~A) \vt~ from \citet{MonacoVB12}, ~~~~~and B) \vt~= 1.3 \kms. The symbols are the same as in Fig. \ref{alfaFe}.  
}
\label{BaFe}
\end{figure}

The cluster M4 is known to be rich in neutron-capture elements like Sr and Ba.
For our sample of turnoff and subgiant stars we found  that, if we adopt the microturbulence velocity of \citet{MonacoVB12}, the mean value of [Ba/Fe] is equal to $\rm +0.33 \pm 0.13$ (Fig. \ref{BaFe}A), and $\rm +0.45 \pm 0.13$ with \vt~=1.3 \kms (Fig. \ref{BaFe}B) .

These values are in good agreement with 
\citet{MarinoVP08} who found in giants below the bump $\rm [Ba/Fe]=+0.41$.
\citet{IvansSK99}  found a higher value of the abundance of Ba : ([Ba/Fe] = 0.60 \,dex $\pm 0.10$) but they studied a more evolved sample of giants and the NLTE correction can be higher for these stars. 

We interpolated in the grid of \citet{KorotinAH15} the NLTE corrections of the Ba abundance for our sample of turnoff stars and for a typical evolved giant. The mean correction is about $\rm-0.05\,dex$ for turnoff stars and $\rm-0.15\,dex$ for giants. Taking these corrections into account the main [Ba/Fe] value in our sample of turnoff stars is +0.28\,dex, or +0.40 with \vt~=1.3 \kms~ and 
it is +0.45\,dex in the sample of giant stars studied by \citet{IvansSK99}.
These values are compatible within the errors.

In our sample of stars we measured  $\rm [Sr/Fe]=+0.37 \pm 0.13$ adopting the microturbulence velocity of \citet{MonacoVB12}, and $\rm [Sr/Fe]=+0.41 \pm 0.13$ with \vt~=1.3 \kms. 
\citet{YongKL08} found a significantly higher value of the abundance of Sr ([Sr/Fe] = 0.73 \,dex $\pm 0.10$) in their sample of giants. But here again the difference can reflect the fact that the results of the LTE computations of the abundance of Ba are not affected in the same way by the NLTE effects when stars are dwarfs or giants. However our measurement of the strontium abundance is based on the profile of a complex feature (see Fig. \ref{Sr2line}) and should be affected by neighbouring lines.

\begin{figure}[h]
\resizebox{\hsize}{!}                   
{\includegraphics {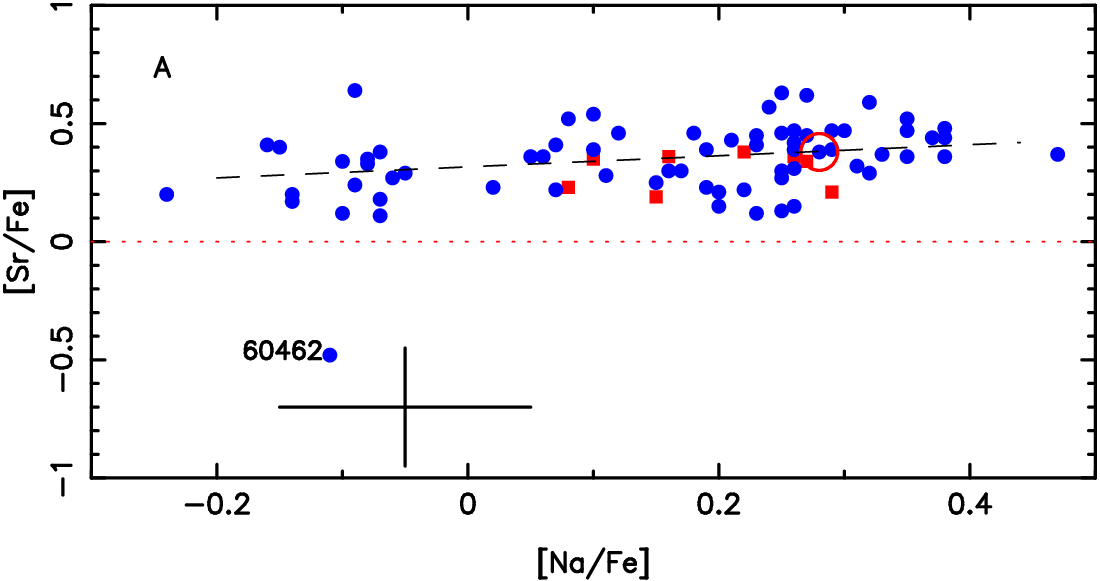}}
\caption[]{Relation between [Sr/Fe] and [Na/Fe]. The symbols are the same as in Fig. \ref{alfaFe}.  
}
\label {SrFe}
\end{figure}

No significant trend is observed between the abundance of Na and the abundances of Sr or Ba (Fig. \ref{BaFe} and \ref{SrFe}). However a difference of 0.2dex in [Sr/Fe] between the Na-poor stars and the Na-rich stars, like the difference observed for Y in the giants by \citet{VillanovaG11} (their Figure 5, lower panel), cannot be excluded.
One star M4-60462 seems to have a weak strontium feature but the abundance of barium is normal in this star. One more spectrum would be useful to confirm this anomaly.

\subsection {Lithium abundance vs. [Na/Fe] and [C/Fe]}

\subsubsection {Subgiant stars}
\begin{figure}[h]
\resizebox{\hsize}{!}                   
{\includegraphics {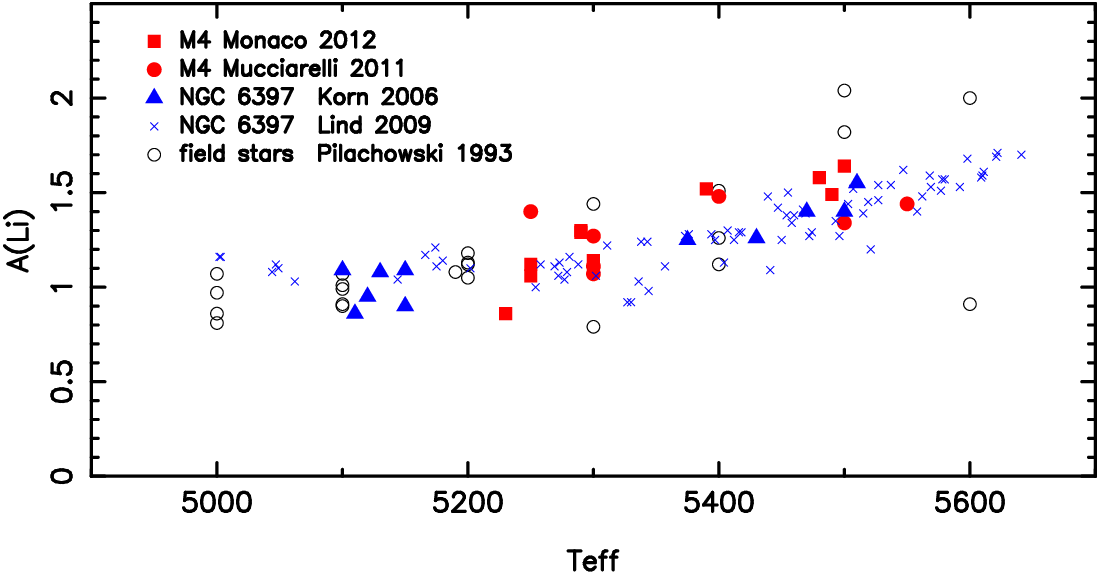}}
\caption[]{Variation of the lithium abundance in evolved stars in M4, NGC\,6397, and field metal-poor stars as a function of the effective temperature. 
}
\label {lisub}
\end{figure}

Low-mass stars leaving the main sequence develop surface convection zones, which deepen when the star continues to evolve. The surface convection mixes the surface layer with deeper material in which lithium has been depleted. 
\citet{MonacoVB12} and \citet{MucciarelliSL11} have shown that the Li abundance decreases in M4 subgiant stars with decreasing stellar effective temperature. This decline of A(Li) (Fig. \ref{lisub}) is the same in NGC\,6397 \citep{KornGR07} and in field stars with about the same metallicity as M4 \citep[see e.g.][]{PilachowskiSB93}.
The combined effects of diffusion and convection are expected to explain the decrease of A(Li) but \citet{MucciarelliSL11} were not able to reproduce the general behaviour of Li and Fe with \Teff~ in M4 (from turnoff stars to giants) starting with the cosmological lithium abundance predicted by WMAP+BBNS. 

\subsubsection {Turnoff stars}

\begin{figure}[h]
\resizebox{\hsize}{!}                   
{\includegraphics {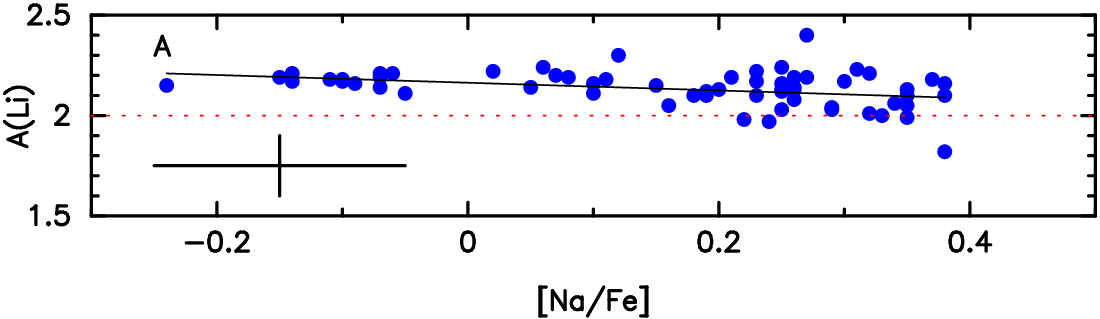}}
\resizebox{\hsize}{!}                   
{\includegraphics {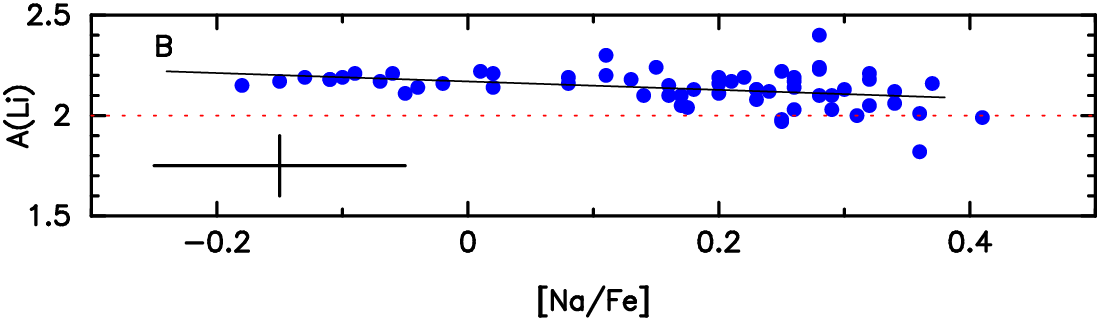}}
\caption[]{A(Li) vs. [Na/Fe] for turnoff stars selected in our sample within a very small interval of temperature: $\rm 5880 \leq T_{eff} \leq 5980 K$ and with ~~~~~   A) \vt~ from \citet{MonacoVB12}, ~~~~~and B) \vt~= 1.3 \kms.}
\label {linaturnoff}
\end{figure}

\begin{figure}[h]
\resizebox{\hsize}{!}                   
{\includegraphics {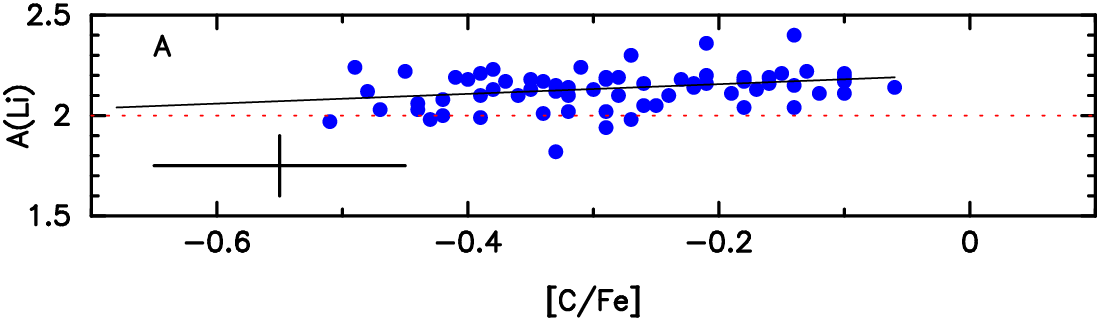}}
\resizebox{\hsize}{!}                   
{\includegraphics {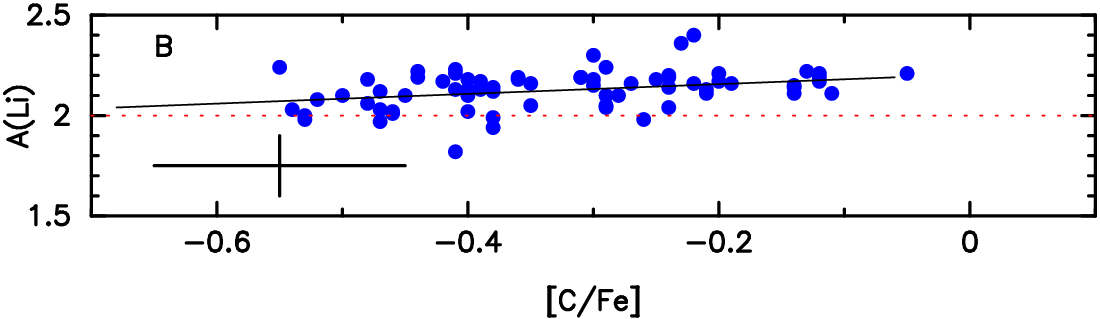}}
\caption[]{A(Li) vs. [C/Fe] for turnoff stars selected in our sample within a very small interval of temperature: $\rm 5880 \leq T_{eff} \leq 5980 K$ and with ~~~~~   A) \vt~ from \citet{MonacoVB12}, ~~~~~and B) \vt~= 1.3 \kms.}
\label {licturnoff}
\end{figure}

In Figs. \ref{linaturnoff} and \ref{licturnoff} we selected, as in \citet{MonacoVB12}, only stars in the temperature interval $\rm 5880 \leq T_{eff} \leq 5980 K$ and we excluded the peculiar Li-rich star M4-37934 (70 stars are taken into account).
 
It has been shown (in Fig. \ref{linaturnoff}  and \citet{MonacoVB12}) that the cluster stars in this sample display a mild but statistically significant Li-Na anticorrelation, at variance with the stars in NGC\,6752 \citep{PasquiniBM05} where the anticorrelation is strong, and the turnoff stars in NGC\,6397 \citep{GonzalezBC09,LindPC09} where no anticorrelation is observed.

When the lithium abundance is plotted versus [C/Fe] (Fig. \ref{licturnoff}) for the same sample of stars, a mild correlation is observed (correlation coefficient +0.30): A(Li) increases slightly with [C/Fe]. This behaviour is expected since the abundances of C and Na are anticorrelated (Fig. \ref{CFe}). 

A change of the microturbulence velocity (panels A and B of Figs. \ref{linaturnoff}  and \ref{licturnoff}) does not induce a significant change in the relations A(Li) vs. [Na/Fe] and  A(Li) vs. [C/Fe].

\section {Conclusions}

The abundances of C, Si, Ca, Ni, Sr, Ba have been measured in a sample of M4 subgiant and turnoff stars, previously analysed for Li by \citet{MonacoVB12}.

\begin{enumerate}

\item
In this paper all the abundance determinations have been done with two options for the microturbulence velocity. In the first option (A) we adopted the same  microturbulence velocity as in \citet{MonacoVB12} and in the second (B) we adopted  \vt= 1.3 \kms~ for all the turnoff and subgiant stars.
With the hypothesis A, as in \citet{MonacoVB12}, we found that  [Fe/H] is higher (by 0.1\,dex) in subgiant stars than in turnoff stars. This effect disappears if  a constant microturbulence velocity  is adopted for all the stars of the sample (option B). In this case [Fe/H] is found to be close to $\rm -1.16 \pm 0.06\, dex$ for all the stars.

The relation A(Li) vs. [Na/Fe] and A(Li) vs. [C/Fe] are practically the same with the hypothesese A and B for the microturbulence velocity.
\\ 

\item
 In Fig. \ref{CFe} the distribution of the dots suggest the existence of two subgroups of stars, one  ``Na-poor C-rich'', and one  ``Na-rich C-poor'', as also noted in giants by \citet{MarinoVP08} and \citet{VillanovaG11} linking this difference to observed photometric differences, recently confirmed by \citet{PiottoMB15}.\\

\item
We confirm the enhancement of the heavy elements Sr and Ba previously found in giant stars. This enhancement has been found to be less important in turnoff or subgiant stars than in highly evolved giant stars \citep{IvansSK99,YongKL08}. 
However, taking into account the NLTE 
 corrections in the computations, the mean Ba abundances become compatible.
 The ratio [Sr/Ba] is found to be almost solar.
The trend between [Sr/Fe] and [Na/Fe] is not significant due to the large errors but a difference of 0.2\,dex in [Sr/Fe] between the Na-poor stars and the Na-rich stars, like the difference observed for Y in the giants by \citet{VillanovaG11}, cannot be excluded.
Additional data about M4 \citep{YongAD14} suggest that at least a part of the heavy elements of M4 could have been produced by Fast Rotating  Massive Stars (known to produce low velocity ejecta supposed to be retained in the cluster).\\

\item
The observed correlations and anti-correlations between Li, C and Na are well represented by the predictions of the model of
\citet{DErcoleAV10}, based on the dilution of ejecta of Super-AGB and massive AGB, which includes sodium and lithium. 
\citet{DAntonaDC12} propose also another interpretation. They show (their Fig. 3) that the dilution of the primordial (cosmological) lithium, at the level of A(Li)=2.7 dex (or even lower), provides an even closer representation of the data of \citet{MonacoVB12}. It may be considered that the production of the elements of the second generation in M4 could be made by fast rotating massive stars (FRMS), that produce sodium but no lithium.  The abundances of Sr and Ba determined in this work are compatible with the abundance pattern produced by FRMS.

It is therefore interesting to look at the abundances of the peculiar, very lithium rich star, M4-37934, observed among the sample of  \citet{MonacoVB12}: this star could be in the middle of the dilution phase \citep{DAntonaDC12}. However, this interesting star has been found similar to the stars of the sample, for all the chemical elements (including Na) that were measurable on the available data. A similar case appeared in a rather similar globular cluster NGC\,6397 \citep{KochLR11}.\\ 

\end{enumerate}

\begin {acknowledgements} 
This work was supported by the ``Programme National de Physique Stellaire'' (CNRS-INSU), and it made use of SIMBAD (CDS, Strasbourg). E.C. acknowledges the FONDATION MERAC for funding her fellowship and A. J. G. acknowledges l'Observatoire de Paris for funding his fellowship. L.M acknowledges support from "Proyecto interno" of the Universidad Andres Bello. 
\end {acknowledgements}

\bibliographystyle{aa}
{}

\end{document}